# Nanocircuits in loop structures: continuous waves preclude gauge invariant wavelengths

Dr. Arthur Davidson
ECE Department, retired
Carnegie Mellon University,
Pittsburgh, PA 15213, USA
artdav@ece.cmu.edu

**Abstract:** Tunnel junctions for quantum computing require discrete spectra from continuous waves on a doubly connected coordinate or loop. For an electron on a metal ring discrete spectra follow from discontinuous Bloch waves. Can both propositions be true? We find using a gauge function originating in the Lagrangian that continuity on a ring or loop violates gauge invariance of the de Broglie wavelength. This same gauge function shows that Lagrangians for the electron on a ring and the charge on a junction are mutual transforms. Thus persistent current on a metal ring and the Coulomb blockade on a tunnel junction seem to be the same dynamical theory based on discontinuous Bloch waves on the compact perimeter of a circle.

## 1. Introduction

This work aims to remove a dichotomy in nanoscience publications. There are two developed collections of literature that treat the basic quantum mechanics of a particle on a circle differently. One group concerns persistent current flow on metal rings. (Büttiker, Imry, & Landauer, 1983) accepted implicitly the symmetry condition $\psi(s+a) = exp\left(\frac{i}{\hbar}qa\right)\psi(s)$ where *s* is a scalar coordinate around the ring of circumference *a* and *q* is any real value of momentum. This makes *ψ(s)* a Bloch function, which must be considered discontinuous when confined to the perimeter of a circle. The other group concerns Josephson tunnel junctions, where the shape of the configuration space is not directly observable. (Averin, Zorin, & Likharev, 1985) and (Likharev & Zorin, 1985) maintained that the symmetry condition $\psi(s+a) = \psi(s)$ must apply to any wave function on a doubly connected loop or ring. This is the definition of a periodic function on a circle, so that any Bloch phenomena would have to come from some other configuration, such as a simply connected space with a periodic potential. An important result of ours is that a continuous and therefore periodic eigenfunction on a circle intrinsically conflicts with gauge invariance of the de Broglie relation. We therefore support Bloch wave symmetry and challenge the continuity of Schrödinger wave functions on a circle for both groups of literature.

We can see the difference between these two groups with a straight-forward search on-line. Using *Web of Science* we found 1126 direct citations of (Büttiker, Imry and Landauer, 1983), and 411 direct citations of (Likharev and Zorin 1985). We set up a random sample of 50 papers from the 411 and manually examined their references. We found 2 papers that also had a reference to *Büttiker et al.* Six of the 50 papers were unavailable to read, so we may estimate that roughly 4 per cent of the set that references *Likharev et al* also references *Büttiker et al.* and also that less than two percent of papers referencing *Büttiker* also reference *Likharev*. This is not proof, but it is consistent with the separation of the two groups more or less by use of different theories. Our position here is that both groups should be covered by a single theory: Bloch waves on loops.

One reason that the general reader of both groups might miss the difference is that both groups rely on Bloch waves. Generally, for the papers we have read carefully in the group established by *Likharev et al,* Bloch waves are used on simply connected spaces with a periodic potential energy. For example, (Averin, Zorin, & Likharev, 1985) falls into this pattern. In contrast papers from the other group often allow Bloch waves on circles, rings, and loops with or without potentials, such as a *Science* perspectives article (Birge, 2009).

We start the analysis below by deriving 1D Lagrangians for an electron on a ring and also for a current through a tunnel junction. We will show by changing variables that these two Lagrangians are gauge transforms of each other (*see eq. 3*). Then we quantize the momentum of the electron on a ring, and we show how the Lagrangian gauge term interferes with gauge invariance (*see eq. 8*). We could use the same mathematics for the tunnel junction. We are left with Bloch waves, which satisfy everything except continuity: they are differentiable; they enable separation of variables in Schrödinger's equation (*see eq. 10* ); they contain the classical solution as a limit (*see eq. 14*); they maintain a discrete spectrum of eigenvalues implied by the compact space of a circle; they satisfy Noether's limit (Byers, 1998) of a single conserved quantity for a particle on a circle; and they explain the normal-electron to Cooper-pair transition (*e to 2e*) as the temperature of the tunnel junction goes below the critical temperature for superconductivity *Tc*. (*see eq. 17*). Their probability density and current density retain continuity on a circle.

## 2. Two Lagrangians are mutual gauge transforms

We imagine a single electron and an applied electromagnetic (EM) field filling the same simply connected 3D space. Then the EM gauge function $\chi(\vec{r}, t)$ (Jackson & Okun, 2001) must be continuous everywhere, and its gradient will have no circulation. We thus preserve the value of the gauge invariant magnetic flux through the region. Because we define both the particle and the field at every point in the simply connected configuration, we need only a single gauge function. We accept the following Lagrangian (Landau & Lifshitz, 1966):

$$L_{3D} = \frac{m(\vec{v} \cdot \vec{v})}{2} - e_0\left(V(\vec{r},t) - \vec{v} \cdot \vec{A}(\vec{r},t)\right) + \frac{d\chi(\vec{r},t)}{dt}. \tag{1}$$

$\vec{r}$ is the position of the particle, and also the coordinate for the scalar and vector potentials $V(\vec{r},t)$ and $\vec{A}(\vec{r},t)$. The particle's velocity is $\vec{v}$. If $V(\vec{r},t)$ is defined at every point in configuration space it must be conservative, so any non-conservative force will be embodied in the $\vec{v} \cdot \vec{A}$ term.

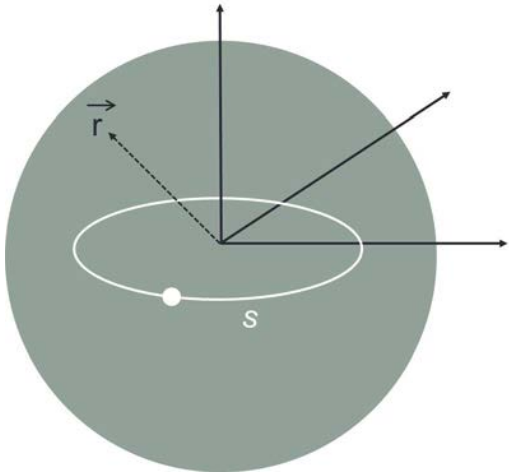

**Fig. 1**. The particle is restricted to a 1D circle *s*, which is doubly connected. An electromagnetic field is configured on simply connected 3D space.

Now without changing the EM field we shrink the path of the particle to the circle *s* in Fig. 1, the situation for Büttiker, Imry and Landauer. The configuration for the particle is now a twice-connected 1D line with circular symmetry different from the 3D simply connected domain of the EM field. If we apply a uniform magnetic field $\vec{B} = \vec{\nabla} x \vec{A}$ perpendicular to the plane of the ring then the particle's velocity vector will be parallel to the circle for all *s*. In this uniform case $\vec{A}$ can also be assumed parallel to $\vec{v}$ and spatially constant along *s*. Then all the vector variables in the 3D Lagrangian become scalar functions of *s*. The vector potential on s is the scalar *A(t)*. The velocity vector becomes the scalar $\dot{s}$. (We use interchangeably a dot over a variable or the Latin *d* to indicate a full time derivative. For partials we always use the Greek ∂.) This leads to our 1D Lagrangian:

$$L_{1D} = \frac{m}{2}\dot{s}^2 - e_0 V(s,t) + e_0 \dot{s} A + \left.\frac{d\chi(\vec{r},t)}{dt}\right|_s + \frac{d\xi(s,t)}{dt}. \quad (2)$$

Eq(2) has two gauge functions with different symmetries due to the different configuration spaces of the field and particle. **The $\chi$ gauge function is** still for the 3D simply connected EM field evaluated at *s* as indicated, and can have no circulation on the ring.

**The particle gauge function $\xi$** in (2) is defined only on the 1D circle. Since we have not defined $\xi$ over the surface, we cannot apply Stokes' theorem. Thus $\xi$ *(t, s)* can have no effect on the magnetic field regardless of its functional form: ∂$\xi$ *(t, s)*/∂s *could have finite circulation without changing the magnetic field.* (Byers & Yang, 1961) and (Wilczek, 1982) used the same type of particle gauge function with a circulation, although they maintained EM terminology.

Now we derive the Lagrangian $L_C$ for a tunnel junction by assuming a lumped capacitance *C* being charged by a current $I = \dot{Q}$. The scalar voltage *v(t)* is the generalized velocity. The coordinate will be a scalar $\phi = \int v(t)dt$. Let the particle gauge function be $\xi(\phi, t)$ similar to Eq.(2). Our circuit Lagrangian $L_C$ will be:

$$L_C = \frac{C\dot{\phi}^2}{2} - V(\phi, t) - \dot{Q}\phi + \frac{d\xi(\phi,t)}{dt}. \quad (3)$$

We show here that Lagrangians (2) and (3), despite three obvious differences, are equivalent: they produce the same classical and quantum equations of motion. Trivially, (2) is constructed in the terminology of real 3D space, where (3) is in lumped circuit space. Because (2) is a ring embedded in real space, it requires the two gauge functions with different symmetries for the 3D field and the circular path for the electron. We derived Eq. (3) entirely in 1 D lumped circuit space where a magnetic field cannot exist, so that the particle gauge function is the only one necessary.

We suppose that the term $e_0 \dot{s} A$ in (2) should correspond to $\dot{\phi} Q$ in (3), since $\dot{s}$ and $\dot{\phi}$ are the corresponding velocities. However what appears in (3) is $\dot{Q}\phi$. We resolve this issue using the arbitrariness of the gauge function to set $\xi(t,s) = -e_0 A s$ on the circle in (2), so that $\frac{d\xi(t,s)}{dt} = -e_0(\dot{A}s+\dot{s}A)$. We thus gauge transform the non-conservative Lagrangian term into $-e_0 \dot{A} s$, the same form as in Lagrangian (3), namely, $-\dot{Q}\phi$. Since $s = 0$ and $s = a$ are the same spatial point, this generalized potential is double valued there. This is allowed because this term is differentiable with a boundary condition, and we are constructing a *non*-conservative force. *So our dynamical theory works for the momentum of an electron on a ring in a uniform time dependent magnetic field <u>and</u> for the charge on a tunnel junction driven by a finite current.*

### 3. Resolving Gauge Invariance

Now we aim to show that if we impose continuity on a momentum eigenfunction on a metal ring then the eigenvalue can retain gauge invariance while the de Broglie wavelength does not. That is, the eigenfunction's wavelength will depend on the arbitrary choice of $\xi(t,s)$. Others have not done this probably because they never realized that the particle gauge function $\xi(t,s)$ can be distinct from the EM gauge function $\chi(t,r)$.

We start quantization with the classical **canonical momentum** $p_c$:

$$p_c \equiv \frac{\partial L_{1D}}{\partial \dot{s}} = p_k + e_0 A + \frac{\partial \xi}{\partial s} \tag{4}$$

We let $p_k \equiv m\dot{s}$ be the kinetic momentum, and for simplicity we assume $\chi$ resides implicitly in $A$. We solve (4) for $p_k$, replacing $p_c$ with the differential operator to get **Schrödinger's momentum operator $\hat{p}$**:

$$\hat{p} = -i\hbar \frac{\partial}{\partial s} - e_0 A - \frac{\partial \xi}{\partial s} \tag{5}$$

We write the momentum eigenfunction $\psi_k$:

$$\psi_k = \Gamma e^{\frac{i}{\hbar}(p_k s + e_0 A s + \xi)} \tag{6}$$

where $\Gamma$ is a normalizing constant. The eigenfunction equation is simple because the $A$ and $\xi$ terms cancel in the pre-factor, leaving the eigenvalue as $p_k$:

$$\hat{p}\,\psi_k = p_k\,\psi_k \tag{7}$$

*Eq. (7) establishes for us that $p_k$ is the gauge invariant eigenvalue for $\psi_k$.* Now we use the *Averin, Zorin* and *Likharev* symmetry condition, $\psi_k(s+a) = \psi_k(s)$ to find the allowed values of $p_k$. There will be an integer number $z$ of wavelengths $\lambda$ in the circumference $a$, or $\lambda = a/z$. We write this in terms of the phase in (6) as $(p_k + e_0 A)a + \Delta\xi = zh$ where $\Delta\xi = \xi(a) - \xi(0)$. In terms of the wavelength $\lambda$ we get:

$$(p_k + e_0 A)\lambda = h - \frac{\Delta\xi}{z}. \tag{8}$$

Eq. (8) has the correct form for the de Broglie relation if $\Delta\xi \equiv 0$, or $\xi$ is continuous on the circle. However, the other term in the phase $(p_k+e_0A)s$ is necessarily discontinuous for finite $p_k$, so we must allow a similar discontinuity in $\xi$. Since we wrote (8) assuming $\psi_k$ is continuous, we must have $(p_k+e_0A)a$ as an integer multiple of $h$. Also $\Delta\xi$ should be an integer multiple of $h$ for the same reason. This means $\Delta\xi$ is finite in (8), and *therefore Eq.(8) violates the gauge invariance expected for the de Broglie relation. On a circle, we can have a continuous eigenfunction OR we* *can have a gauge invariant wave length, but not both.* A continuous eigenfunction and the de Broglie wave length may both be regarded as fundamental, but a wavelength that depends on gauge choice is not a solution at all. Therefore we discard continuity and show below in Section 4 that Schrödinger's equation remains intact.

We still have two conditions on the momentum eigenfunction: the compact nature of the circle still implies a discrete spectrum, and Noether's theorem (Byers, 1998) still allows only a single variable to be conserved. These conditions are met by a Bloch momentum eigenvalue $p_k = q + \frac{zh}{a}$. $p_k$ is the single conserved quantity that we can distribute arbitrarily over the terms $q$ and $zh/a$, where $q$ is real and $z$ is an integer. If we apply an external force momentum is not conserved and q can respond smoothly. For an individual eigenfunction we can easily transfer accumulated value to the integer z, as usual for a reduced zone on a band diagram.

The Bloch form in superposition also maintains the discrete spectrum we need. If each superposed eigenfunction has the same value of $q$ our superposition will be:

$$S = e^{\frac{i}{\hbar}(qs)} \sum_z b_z e^{\frac{i2\pi zs}{a}} \tag{9}$$

where $b_z$ is a complex coefficient. By Fourier's theorem the summation in (9) is periodic over the circumference *a* so *S* retains the Bloch wave symmetry of each of its component waves. This implies that all waves on a ring will be of the Bloch form. The point is that (9) is a discrete spectrum; we have merely shifted the momentum of each term in the sum by $q$ due to the common factor $e^{\frac{i}{\hbar}(qs)}$.

The Bloch wave *S* above would have to be differentiable despite a singularity in the pre-factor $e^{\frac{i}{\hbar}(qs)}$ at *s=0* and *s=a*. The boundary condition is that the constant slope of the phase is the momentum eigenvalue, which must be continuous across the singularity. <u>Therefore we can differentiate a discontinuous eigenfunction on a ring if its discontinuity is a jump in its phase, as for $e^{\frac{i}{\hbar}qs}$</u>.

## 4. Schrödinger's equation in real space

We show that Bloch waves allow the separation of continuous and discrete solutions of Schrodinger's equation. The quantum Hamiltonian from $L_{1D}$ and $V=0$ is

$$\boldsymbol{H} = \frac{1}{2m}\left(-i\hbar\frac{\partial}{\partial s} - e_0 A - \frac{\partial \xi}{\partial s}\right)^2 - \frac{\partial \xi}{\partial t}. \tag{10}$$

It turns out that the momentum eigenfunction (6) also solves Schrödinger's equation given $\boldsymbol{H}$ above, with an added term for the total energy $E(t)$:

$$\left[\boldsymbol{H} - i\hbar\frac{\partial}{\partial t}\right]\Gamma exp\left(\frac{i}{\hbar}\left(\left(\frac{zh}{a} - q(t) + e_0 A(t)\right)s + \xi(s,t) - \int E dt\right)\right) = 0 \tag{11}$$

We take the $s$ and $t$ partials and cancel the wave function from each term. Then these 5 terms must add to zero:

$$\frac{1}{2m}\left(\frac{zh}{a} - q\right)^2 + \left(\frac{\partial \xi}{\partial t} - \frac{\partial \xi}{\partial t}\right) - \dot{q}s + e_0 \dot{A}s - E(t) = 0 \tag{12}$$

*We cancel the terms with $\xi$ as required by gauge invariance*. We sum separately to zero all the terms dependent on $s$, leaving the results valid for all s:

$$-\dot{q} + e_0 \dot{A} = 0 \tag{13}$$

and

$$E(t) = \frac{1}{2m}\left(\frac{zh}{a} - q(t)\right)^2. \tag{14}$$

As required by the correspondence principle if $h$ approaches zero (13) and (14) approach the classical results. We re-write (13) and (14) independent of the EM gauge by doing a line integral around the ring to get $E(t) = \frac{e_0^2}{2ma^2}(z\phi_N - \Phi(t))^2$, where $\Phi$ is the magnetic flux through the ring. This equation is equivalent to the results for a single electron on a ring derived heuristically in (Büttiker, Imry, & Landauer, 1983). *Nota bene:* Averin, Likharev and Zorin do not obtain the known classical result. For them *E→0* as *h→0*.

## 5. Schrödinger's equation in lumped circuit space

We turn to apply these results written in terms of a system in real space to the corresponding system in 1D lumped circuit space. This is do-able because these two systems share equivalent configurations and Lagrangians. We can transcribe results: *m* → *C*, *s* → *ϕ*, *e₀A* → *Q*. *Q* is a generalized momentum, and *Qϕ* has the units of *h*. If we assume the generalized circumference of the ring to be $\phi_N = h/e_0$ we will get quantized charge as discrete electrons. (We reserve $\phi_0$ for the superconducting flux quantum $\phi_0 = h/2e_0$, which will later be the period of the conservative potential *V(ϕ,t)*.)

If *V(ϕ,t) = 0* we get $C\ddot{\phi} = -\dot{Q}$ as the Euler-Lagrange equation of motion. This is the standard classical linear circuit result independent of the $\xi$ gauge function. On a ring $\dot{Q}\phi$ has a point singularity, but we can differentiate anyway with the use of a physical boundary condition that the force is continuous. By transcription from (13) and (14) the energy band equations for the tunnel junction are:

$$\dot{Q} = I \tag{15}$$

$$\frac{1}{2C}\left(\frac{zh}{\phi_N} - q(t)\right)^2 = \frac{1}{2C}(ze_0 - Q(t))^2 = E(t) \tag{16}$$

Equations (15) and (16) are an exact solution for a tunneling capacitor without Josephson coupling. We plot them in **Figure 2** below. The vertical axis is the square root of the energy normalized to $e_0^2/2C$ so that parabolic segments of Equations Eq. (16) become the solid diagonal straight lines. We follow the Bloch oscillation path along a-b-c-d-e-a. Integration starts at $t = 0$ along the $z = 0$ curve at point a where $Q/eo= 0$. The system moves along the dotted line in the ground state from point a to point b where the total charge is about $+e_0/2$. "Coulomb blockade" refers to the blockage of tunneling events because the system is already in the ground state at b, and cannot get to c due to de-coherence by the environment. A transition to c keeps total charge *(ze$_o$ - Q)* constant as *Q* and *z* change by – and + *e$_o$*. State c integrates to d along the *z = -1* curve, and then the system tunnels diabatically to point e, reducing the total charge by e$_o$ back to zero, the initial charge state.

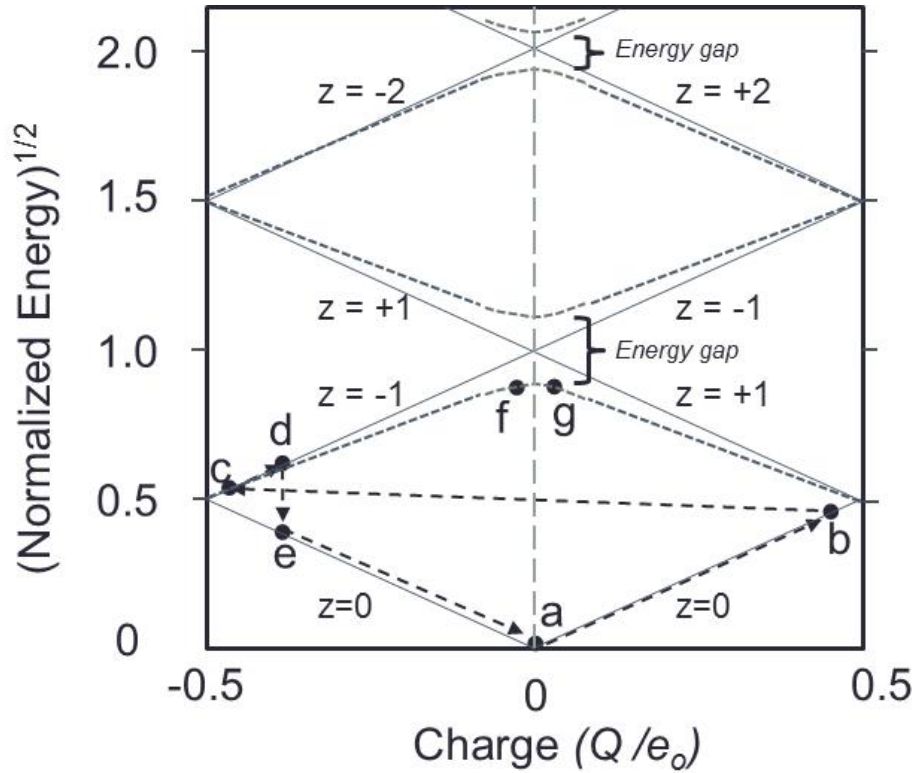

**Figure 2.** Square root of energy versus net charge through a tunnel junction. Solid crossed lines are for no Josephson effect and form a graph of eq. (17). Dashed lines *a* through *e* show a single particle Bloch oscillation. Dotted lines show developing energy gaps for small Josephson coupling. Transition from *f* to *g* is adiabatic changing the charge by a Cooper pair.

*The moments of inertia,* ~*ma$^2$* for the ring and ~*C$\phi_N^2$* for the tunnel junction, are related by fundamental constants so that measurements of the two systems can be compared. The ring moment-of-inertia *ma$^2$* and the tunnel junction generalized moment of inertia *C$\phi_N^2$* share the same units, which can be expressed as Joule-second$^2$. These two moments will be equal when $\frac{a^2}{C} = \frac{\phi_N^2}{m} \cong 18.8$ in mks units. Thus e.g. a 10 femto-farad tunnel junction corresponds to a metal ring with a circumference of 0.43 microns.

*Nota bene:* An important result of our gauge invariant theory is that a single electrically isolated current driven normal tunnel junction can account for the Coulomb blockade behavior expected for single electrons. The single electron transistor (Stewart Jr. & Zimmerman, 2016) with a single-electron box between two tunnel junctions is not the minimum apparatus necessary. In fact, not having the isolated box may alleviate problems of randomly induced charge in the operation of these devices.

**We now turn on the internal Josephson potential energy,** with a period *$\phi_0$* half that of the circumference *$\phi_N$* of the ring. Then there is one dynamical equation that covers single electron dynamics above the superconducting transition *T$_c$*, *and* Cooper pair dynamics added below *T$_c$*. This model should not be confused with the single electron transistor (SET) where measurements of current have shown a transition from single electrons to pairs. (Tinkham, 1996). Indeed, there appear to be few measurements above and below T$_c$ for an isolated junction, in contrast to the case for the SET. The reason for this may be that papers like (Averin & Likharev, 1986) or (Watanabe & Haviland, 2003) allow only a simply connected coordinate: they predict no Bloch phenomena above Tc where the periodic Josephson potential vanishes. The present work supports the idea of measuring the same single junctions with adequate isolation above and below $T_c$. Observation of *e.g.* the single electron Coulomb blockade in single tunnel junctions above Tc would be dispositive for the generalized coordinate to have a ring shape.

In the Lagrangian Eq. (3) below $T_c$ we let $V(\phi,t) \to E_J\cos(\theta)$ where $E_J$ is the temperature dependent Josephson potential energy coefficient and $\theta = (2\pi\phi/\phi_0)$ is the Josephson angle. For simplicity we take $\xi = 0$, and let $E_j$ be small. We get the following forms for the Hamiltonian and wave function

$$\mathcal{H} = \frac{1}{2C}\left(-i\hbar\frac{\partial}{\partial \phi}\right)^2 - E_J\cos\theta + \dot{Q}\phi \quad and \quad \psi = e^{\frac{i}{\hbar}iq\phi}U(\phi,t) \tag{17}$$

The complex function $U(\phi,t)$ is the periodic part of the Bloch wave we are looking for. Qualitative solutions are shown for (17) as the dotted lines in Fig. 2. An energy gap opens up at $Q = 0$, but not at the band edge $Q = \pm 0.5\ e_0$. To see this consider the following integral:

$$(Energy\ gap: q \to 0) \propto \int_{-2\pi}^{2\pi} \cos(\theta)\left(\cos^2\left(\frac{\theta}{2}\right) - \sin^2\left(\frac{\theta}{2}\right)\right)d\theta = 2\pi \tag{18}$$

The *cos(θ)* factor is the Josephson potential energy. The squared terms are approximations of the probability density for waves above and beneath the gap. At the zone edge where $q = e_o/2$ the arguments for the squared terms in (18) become not *θ/2* but *θ/4*, and the gap vanishes along with the definite integral. This means that transitions like from d to e remain dissipative single electron events. But a qualitatively new transition opens up below the gap from f to g, changing charge by two electrons without changing energy (Büttiker, 1987). When normalized we get this energy gap quantitatively equal to $E_j$. For high enough current $I$ the system will have occasional **Zener tunneling** across the gap to the higher bands at $Q/e_o = 0$.

## 6. Recapitulation

There is a highly regarded 1983 paper by Büttiker, Imry, and Landauer where the ring geometry is an experimental fact, and the authors predict persistent currents by assuming implicitly that the wave functions are discontinuous Bloch waves. There are two other highly regarded papers from 1985 by Averin, Zorin and Likharev where the ring geometry must be inferred. These authors restrict Bloch phenomena to simply connected domains with periodic potentials (that is, not the ring geometry) by invoking continuity of all wave functions on a ring. We demonstrated here that imposing continuity of wave functions on any ring leads to an unacceptable absence of gauge invariance of the de Broglie wavelength.

The set of superposable Bloch wave eigenfunctions on any ring is complete allowing solutions of Schrödinger's equation to be separated into gauge invariant discrete and continuous parts. The discrete Bloch wave eigenfunctions are discontinuous but differentiable. We predict that all quantum waves on a ring are Bloch waves, vindicating Büttiker, Imry and Landauer. Any system modeled as a particle on a 1D ring will have equivalent dynamics despite different terminology (Davidson & Santhanam, 1990). A tunnel junction and a metal ring with an electron can have comparable moments of inertia, so experimental results can be compared directly.

By carefully working out the behavior of two gauge functions with different properties, we have shown the dynamical equivalence of an electron moving on a metal ring and a charged Josephson tunnel junction with a generalized circular configuration. Their Lagrangians are mutual gauge transforms. Bloch waves remain on a 1D ring even without a periodic potential.

The Bloch momentum for an electron and the corresponding charge for a tunnel junction can be changed continuously by the external non-conservative force, while the system makes discrete transitions between bands. Uniquely, our quantum solutions approach the classical limit as *h* goes to zero. The full device structure of a single electron transistor should not be needed to see the Coulomb blockade effect: Our theory implies that a single-electron box is not needed. Reduction of charge drift may be a consequence. (Stewart Jr. & Zimmerman, 2016)

To more fully understand qubit tunnel junction dynamics (Amin, 2005) (Devoret, Wallraff, & Martinis, 2008) the more complete gauge invariant model may be necessary: when a Josephson tunnel junction is above its critical temperature there is no periodic potential, but the periodicity of the ring maintains single electron charging behavior. As T falls below $T_c$, the Josephson periodic potential grows, leading to energy gaps and adiabatic paired electron charging phenomena. Finally, (Blackburn, Cirillo, & Groenbeck-Jensen, 2016) showed the importance of comparing experiments to both quantum and classical theories.

## Acknowledgments

Dr. Alan Kadin of Princeton Junction New Jersey Prof. N. F. Pedersen of the Technical University of Denmark, Dr. Mohammad Amin of D-Wave Systems and Dr. Alan Kleinsasser of the Jet Propulsion Laboratory provided encouragement and critical readings of manuscripts.